\documentclass[11pt,a4paper]{article}
\usepackage[hyperref]{ranlp2021}
\usepackage{times}
\usepackage{latexsym}
\usepackage{graphicx}

\usepackage{microtype}

\aclfinalcopy 
\def\aclpaperid{222} 

\title{A Domain-Independent Holistic Approach to Deception Detection}

\author{Sadat Shahriar \\
  University of Houston, \\
  Texas, USA\\
  \texttt{sshahriar@uh.edu} \\\And
  Arjun Mukherjee \\
  University of Houston, \\
  Texas, USA\\
  \texttt{arjun@cs.uh.edu} \\ \And
  Omprakash Gnawali \\
  University of Houston, \\
  Texas, USA\\
  \texttt{gnawali@cs.uh.edu}
  }

\date{}

\begin{document}
\maketitle
\begin{abstract}
The deception in the text can be of different forms in different domains, including fake news, rumor tweets, and spam emails. Irrespective of the domain, the main intent of the deceptive text is to deceit the reader. Although domain-specific deception detection exists, domain-independent deception detection can provide a holistic picture, which can be crucial to understand how deception occurs in the text. In this paper, we detect deception in a domain-independent setting using deep learning architectures. Our method outperforms the State-of-the-Art (SOTA) performance of most benchmark datasets with an overall accuracy of 93.42\% and F1-Score of 93.22\%. The domain-independent training allows us to capture subtler nuances of deceptive writing style. Furthermore, we analyze how much in-domain data may be helpful to accurately detect deception, especially for the cases where data may not be readily available to train. Our results and analysis indicate that there may be a universal pattern of deception lying in-between the text independent of the domain, which can create a novel area of research and open up new avenues in the field of deception detection. 
\end{abstract}

\section{Introduction}

In the current era of the flood of information, deception has become an undeniable event, causing a financial or political catastrophe and even the loss of human lives. Often, we do not have the necessary resources to validate a tweet or a catchy social media forwarded news link. Our idea is to capture the writing style hidden ``between-the-lines'', intended to deceive the reader. The deception can be in the form of any textual stream and on any topic. So, our objective is to find a holistic model that can leverage thousands of textual resources and find a learning architecture to decode the deception. 

Adopting the definition of \textit{Deception} from Burgoon and Buller, we define \textit{Deceptive Text} as any textual content that aims to misconstrue an affair in a  deliberate way causing the reader at a disadvantage either directly or indirectly \cite{Burgoon1994}. Deceptive text can be of various forms. For example, in the news and public media domain, the deceptive text is known as Fake News. In social media, a deceptive text can happen in the form of a rumor. Spam or a phishing email is treated as deceptive content in the personal mail or messaging domain. Each domain's deceptive text has distinct ways to deceit the reader. While fake news can spread falsified propaganda, spam or phishing email can be used for merely monetary gain. Therefore, the ways of formulating a deceptive text can have significant variations. Notwithstanding, all deceptive texts share a common goal of tricking the reader and thus, a general deception pattern should exist in these texts. Unravelling the pattern can play a pivotal role to provide a holistic view of deception, which in-turn can bolster the deception-detection. However,  Gr\"ondahl and Asokan indicated that existing works fail to generalize the deception across different domains \cite{Grondahl_survey}. In this work, we hypothesize that-- \textbf{(H1)} A deep learning architecture trained on a generalized deceptive-writing setting (both in-domain and out-of-domain data) can better understand the underlying general pattern of deception than using the in-domain data only.

On the rise of a fairly new event, we may not have data available to detect deceptive text floating around social and mass media, especially the labeled data for supervised training. Such occasions pose a unique challenge to stop the spread of misinformation. A holistic deception detection system can come in handy in such situations. For example, although the pandemic was as ancient as human civilization, in the age of massive data availability, COVID-19 becomes a new event, and misinformation caused by this event can be hard to battle. Therefore, we hypothesize that \textbf{(H2)} A generalized dataset can be helpful to detect deception in a new event, even when little or no in-domain data is available. To test our two hypotheses, we train and fine-tune a BERT model, SBERT model, and character-level-CNN model \cite{devlin-etal-2019-bert,reimers-2020-multilingual-sentence-bert}. We further investigate the intermediate-layer learning mechanism using t-SNE visualization and attention-weight analysis.  

Researchers worked meticulously to model deception on different domains, but a unified approach has not been successful. Hern\'andez-Casta\~neda et al. suggested a cross-domain approach for a generalized deception-detection model \cite{Hernandez2017}.  Although they claimed to build a domain-independent system, their choice of the datasets, namely \textit{DeRev}, \textit{OpSpam}, and \textit{Opinion} (a dataset where the participants were told to lie about their opinion about selected topics), are of the type opinion-only. Moreover, the \textit{Opinion} dataset can hardly be treated as deceptive. That is because when people are told to lie about an opinion, the lie may not have the potential of deceiving someone which contradicts our definition of \textit{Deception}. On the contrary, our choice of datasets have at least three variations of categories -- fake news, Twitter rumor, and spam. 

Thus, we summarize the main contributions of this paper as--
\begin{itemize}
    \item Our work is the first to propose a domain-independent holistic approach to detecting deception leveraging available public datasets.
    \item We quantitatively show that for an unseen event, only a fraction of the total available data can be helpful in successfully detecting the deception. 
\end{itemize}

\section{Literature Review}
In this research, we aim to detect deceptive content intended to mislead people rather than entertain them. Therefore, the deceptive contents can be viewed as deceptive news, disinformation, cherry-picking, and click-bait \citep{Zhou}. There have been several approaches for manual fact-checking, both in the form of expert-based \cite{Politifact, gossipcop, factchecker} and crowd-sourced \cite{CREEDBANK}. However, given the enormous influx of information, such manual approaches are time-consuming and often biased. Therefore, automated fact-checking came in handy. Based on how users spread the falsified information, researchers adopted following approaches to detect deceptive content: \textit{news-cascade}, which is a tree-like structure to analyse the propagation in social-media \cite{ma-etal-2018-rumor}, and \textit{Propagation Graph}\cite{zafrani}. However, such approaches are also constrained by the availability of propagation detection resources.
Moreover, false information cannot be detected before it spreads out. Additionally, some research tries to detect false news based on the source credibility \cite{source_cred}. Nevertheless, the stream of new sources now and then makes the task challenging.

Therefore an AI-based method aiming to detect deception based on the textual content can be handy. Because of the fewer dependencies and availability of content, many researchers worked in that direction. Zhou et al. divided the task into two steps: (i) how well the deceptive news content is captured, and (ii) how well the classification model performs to detect deception \cite{Zhou}. Approaches, such as the Bag of Words (BOW) model, POS tagging, rhetorical relationships, were used as features to detect deceptive news \cite{BOW, POS, Zhou}. Nevertheless, we are more interested in the semantics, as the task of deception may lie in between the text. Word level context embeddings, including Continuous Bag of Words (CBOW) and skip-gram models were used to represent text for detecting fake news \cite{DBLP:CBOW}. Along with such representations, several machine learning algorithms are used for classification purposes. Additionally, with the rise of deep learning, Convolutional Neural Network (CNN) and many variants of Recurrent Neural Network (RNN) are used as well \cite{CNN, RNN}. 

Recently the Bidirectional Encoder Representations from Transformers (BERT) model and its variants have gained enormous popularity due to its pretraining capability \cite{BERT}. M\"uller et al. proposed COVID-Twitter-BERT (CT-BERT), which is pretrained on COVID-19 related Twitter messages \cite{covid-twitter_BERT}. Such task-specific BERT model outperforms the generic BERT models in the significant margin in many COVID-related classification tasks, including AAAI2021 COVID-19 shared challenge of COVID-19 Fake News Detection \cite{COVID-19-winner}. A similar fine-tuning approach is used by Liu et al., where they proposed a two-stage approach for short fake news detection. Unlike the original BERT model, they utilized all hidden states to apply the attention mechanism to calculate weights for text representation. Their approach produced a 34\% accuracy in the LIAR dataset \cite{two-stage-LIAR-BERT}. Kaliyar et al. proposed FakeBERT, which uses a CNN model after the BERT embedding layer \cite{kaliyar2021fakebert}. 

Although the current deception detection methods work well, the methods are highly dependent on the training of the specific domain. On the contrary, we intend to eliminate dataset-specific training and train our model for the generic deception detection task.

\section{Dataset}
We curate ten datasets for our analysis. We broadly categorize them as-- i) Spam, ii) Fake News, and iii) Rumour. The details of the datasets are described below.

\subsection{Email and Text Spam}
For the spam datasets, we select two personal-messaging datasets. First, we select \textit{SMS Spam} collection Dataset from UCI Machine Learning Repository \cite{SMS}. This dataset collection has messages collected from different sources totaling 5,574 messages, of which 4,827 are Hams, and 747 are Spams. We also curated the Enron-Spam datasets, which is a benchmark dataset for email spam collection from six different users \cite{Metsis2006SpamFW}. There were 15,421 Spam emails and 14,923 Ham emails, totaling 30,344 emails.

\subsection{Fake News}
For the Fake News datasets, we start by collecting the COVID-19 related fake news. The first one  is \textit{Constraint@AAAI2021 - COVID19 Fake News Detection in English} \cite{patwa2021overview}. The data are collected from various social media platforms. The training data has 6,420 texts, validation data has 2,410 texts, and the test data has 2,140 texts. The dataset has overall 52\% real and 48\% fake news. Another COVID-19 related dataset we use is \textit{Zenodo-- COVID Fake News Dataset} \cite{sumit_banik_2020_4282522}. The Zenodo COVID dataset has 10,201 texts, out of which there are 9,733 fake news and only 468 real news. 

Next, we collect a dataset of varying unreliability, developed by Rashkin et al., where each text was considered as either a \textit{Satire}, a \textit{Hoax} or a \textit{Propaganda} \cite{rashkin-etal-2017-truth}. Unlike our definition of Deceptive-text, Satire cues the reader of the news being a joke only, and thus, we treat Satire as a non-deceptive text. The Hoax and the Propaganda are meant to misguide people, and therefore, we treat them as Deceptive-text. There are 38,859 texts, of which 24,839 were deceptive texts and 14,020 non-deceptive texts. Additionally, they collected 4,362 data from Politifact, which are rated in a 6 pt. scale, namely, True, Mostly-True, Half-True, Mostly-False, False, Pants-on-Fire False. We consider the first three as non-deceptive text, and the last three are deceptive text. The dataset comes with a separate train, test, and a dev set. 

Additionally, we use the \textit{FakeNewsNet} dataset which comes with real and fake news content from PolitiFact and GossipCop \cite{shu2018fakenewsnet, shu2017fake, shu2017exploiting}. In total, there were 23,196 data.

The last dataset we use in the Fake News Category is the \textit{Liar} benchmark dataset \cite{wang-2017-liar}. Along with the text and the labels, the dataset comes with 12 other metadata. The dataset comes with a separate train, test, and a dev set. In total, the dataset contains 12,791 texts.  
\subsection{Rumour}
We collect the PHEME dataset of rumors and non-rumors, which contains the Twitter rumor and non-rumors during breaking news, namely in the events of Charlie Hebdo, Ferguson, Germanwings, Ottawa Shooting, Sydney Siege \cite{Zubiaga2016}. We treat rumors as deceptive text and non-rumors as non-deceptive text. In total, we have 6,425 texts.

\section{Methodology}
\subsection{Deep Learning Frameworks}
\subsubsection{Bidirectional Encoder Representations from Transformers (BERT) model}
BERT is a pre-trained language representation model proposed by Devlin et al. \cite{devlin-etal-2019-bert}. BERT is trained on a bidirectional setting of context and with the objective of Masked Language Modelling and Next Sentence Prediction. The transfer learning capability of BERT makes it a popular candidate for many NLP tasks, such as sentiment classification, fake news detection, question-answering. The BERT model consists of several transformer blocks, which are made of attention and feed-forward layers \cite{vaswani2017attention}. In this work, we fine-tune the \textit{bert-base-uncased} version of the BERT model, which consists of twelve transformer blocks. The 768-dimension output vector from the BERT model (position of $[CLS]$ token) is fed to a one-layer fully-connected network for classification. We use the recommended batch size of 16, and other hyperparameters (epoch, hidden-unit, learning-decay-rate) are chosen by cross-validation. 

\subsubsection{Sentence-BERT}
The Sentence-BERT (SBERT) is a modified version of BERT capable of representing semantically meaningful sentence embedding \cite{reimers-2019-sentence-bert}. SBERT is based on siamese and triplet networks for fine-tuning over BERT. It performs a pooling operation (min, max, or mean pooling) on the output of BERT. SBERT has a much faster running time compared to BERT. In our work, we use the pre-trained SBERT model and fine-tune it with two fully-connected hidden layers on top of that. We use cross-validation to choose hyperparameters (hidden units, batch size, and epochs).
\subsubsection{Character-level-CNN model}
Convolutional Neural Net (CNN) is a popular network of computer vision tasks, and it extends to NLP tasks \cite{kim2014convolutional}. 
The Character-Level CNN (Char-CNN) was first proposed by Zhang et al., which is capable of dealing with Out-Of-Vocabulary (OOV) words by focusing on the character-level rather than the word or sentence level \cite{Zhang2015CharacterlevelCN}. The Char-CNN consists of six convolutional layers and three fully-connected layers, followed by a max-pooling layer. We empirically choose the convolution filters to be 256. The fully-connected layer units, batch size, and the dropout rate is chosen by cross-validation. 
\subsubsection{Ensemble model}
There can be two ways to ensemble the DL models--1. Hard Decision and 2. Soft Decision. In Hard Decision, we make the prediction based on the majority voting on a test sample. However, the majority voting can eliminate the effect of a strong probability confidence model by predicting a class even when two of the three models predict a class with weak probability confidence. Therefore, in the Soft Decision ensemble, we take the average softmax probability score of the DL models before the prediction phase. For illustration, let's assume in a two-class classification setting, the softmax layer output of BERT, SBERT, and Char-CNN is $[b_0, b_1], [s_0, s_1], [c_0, c_1]$ respectively. The ensemble model will have the softmax probability output of, $[\frac{b_0+s_0+c_0}{3}, \frac{b_1+s_1+c_1}{3}]$. In this work, we use the Soft Decision ensemble model.

\subsection{Experimental Set-up}
We use the dataset-provided test set for Liar, Rashkin-Politifact, and  COVID-AAAI datasets for the general holistic deception detection task. For the rest seven datasets, we randomly sample the data into training, validation, and test set as 60\%-20\%-20\% and repeat the experiments for three different splits. We report the average performances of the three splits.

For the new-event holistic deception detection task, we select COVID-19 as a new event and combine the test set of COVID-AAAI and COVID-Zenodo. First, we train only on the out-of-domain eight datasets. Then, we add 20\%, 40\%, 60\%, 80\%, and 100\% in-domain COVID data along with the out-of-domain datasets for training. We also use 20\% of the training data as the validation set. This setup enables us to examine the strength of a holistic model in an unknown or little-known event. 

\section{Results and Discussion}
To test the hypotheses, we divide our experiments in two parts: i) General Holistic Deception Detection and ii) New-Event Deception Detection.

\subsection{General Holistic Deception Detection}
The idea behind a general holistic deception detection task is to build a generalized system that will be able to detect deception in the text irrespective of the topic or target domain. Being a domain-independent system, the holistic model may have more robustness than the domain-specific model.

\begin{table*}[t!]
\small
\begin{center}
\begin{tabular}{c|cc|cc|cc|cc|cc}

\cline{2-11} &
\multicolumn{2}{|c|}{\textbf{Char-CNN}} & \multicolumn{2}{|c|}{\textbf{Sentence-BERT}} &
\multicolumn{2}{|c|}{\textbf{BERT}} &
\multicolumn{2}{|c|}{\textbf{Ensemble}} &
\multicolumn{2}{|c}{\textbf{SOTA}}\\
\hline \textbf{Dataset} & \textbf{Acc(\%)} & \textbf{F1 (\%)}& \textbf{Acc(\%)} & \textbf{F1 (\%)} & \textbf{Acc(\%)} & \textbf{F1 (\%)} & \textbf{Acc(\%)} & \textbf{F1(\%)} & \textbf{Acc(\%)} & \textbf{F1 (\%)}  \\ \hline
PHEME & 80.72 & 81.43 & 83.82 & 78.51 & \textbf{86.41} & 81.72 & 85.21 & \textbf{82.74} & -- & 77.40 \\
Liar & 64.80 & 54.40 & 68.75 & 62.57 & 67.01 & 59.34 & \textbf{70.72} & \textbf{62.60} & 65.54 & 60.80 \\
FNN-Gossipcop & 78.59 & 55.30 & 80.58 & 57.67 &  \textbf{86.11} & {68.10} & 85.69 & 66.70 & {80.80} & \textbf{75.50}  \\
FNN-Politifact & 71.70 & 58.33 & 73.58 & 68.69 & 81.46 & 77.43 & 81.76 & 77.91 &\textbf{90.40} & \textbf{92.80} \\
Rashkin-Politifact & 88.34 & 82.82 & \textbf{95.23} & \textbf{93.46} & 88.62 & 84.92 & 94.66 & 92.46 & -- & 56.00 \\
Rashkin-Newsfiles & 97.81 & 98.25 & 96.42 & 97.15 &  \textbf{99.64} & \textbf{99.71} & 99.43 & 99.56 & -- & -- \\
COVID-Zenodo & 96.04 & 92.55 & 95.78 & 97.77 & \textbf{97.45} & \textbf{98.66} & 97.21 & 98.53 & -- & --\\
COVID-AAAI & 89.39 & 88.67 & 89.62 & 89.03 & \textbf{95.42}  & {95.07} & 95.20 & 94.68 & -- & \textbf{98.37}\\
ENRON email spam & 97.64 & 97.67 & 97.90 & 98.43 & 99.33 & 99.32 & \textbf{99.43} & \textbf{99.46} & 95.88 & 95.76 \\
SMS Spam & 92.82 & 77.78 & 97.12 & 91.40 & 98.32 & 93.56 & \textbf{98.42} & \textbf{94.06} & 97.64 & --  \\
\hline
\textbf{Total} &89.98 & 89.53 & 90.42 & 90.27 & 92.72 & 92.50 & 93.42 & 93.22 & -- & --\\
\hline
\end{tabular}
\end{center}
\caption{ Holistic Deception Detection performance (accuracy and F1-score) with Char-CNN, SBERT, BERT, and Ensemble model. We also present the current SOTA performance for comparison. The bold numbers indicate the best performance on a dataset. The field with ``--'' indicates that the performance is not reported.}
\label{tab:hol:standalone-models}
\end{table*}



From the standalone model performances on Table \ref{tab:hol:standalone-models}, we can see the varying degree of performance across different datasets. However, in almost every case, BERT outperforms the Char-CNN and SBERT model. With the self-attention mechanism, BERT may better capture the nuances within the text than the Char-CNN model, which can give potential cues to detect deception. Despite being a variant of BERT, the SBERT model may lose some information by the fine-tuning and the pooling process, which creates further research direction towards fine-tuning the SBERT model for deception detection. We find the best overall performance for the ensemble model, with an accuracy of 93.42\%, and an F1-Score of 93.22\%. The best performing standalone model -- BERT lags slightly behind that with an accuracy of 92.72\%, and an F1-Score of 92.50\%.   

For the PHEME dataset, we find the best performing F1-Score in the ensemble model as 82.74\%, which is better than the current top scorer stA-HitPLAN based model (77.4\%) \cite{phemeSOTA}. Similarly, for the Liar dataset, the ensemble model achieves the best performance with an accuracy of 70.72\%, and F1-Score of 62.60\%, which outperforms a text-based BERT-CNN architecture by 5.18\% and 1.80\% respectively  \cite{upadhayay2020sentimental}.

The current SOTA F1-Score in FakeNewsNet--Gossipcop (FNN-Gossipcop) and Politifact (FNN-Politifact) is 75.50\% and 92.80\% in F1-Score using a news and user-comment encoder, and co-attention network \cite{dEFEND}. However, their experiments differ from ours in the fact that i) they used the news with at least three user comments, reducing their sample size by 73\% and 60\% compared to the original data size, and ii) besides the news text, they took user comments into account as well. The user comments contain crowd opinion, which provides vital information to detect deceptive content \cite{usercomments, dEFEND}. Nevertheless, selecting the source with user comments can significantly reduce the data space, and thus, we do not use them.

Our SBERT model performs the best in the Rashkin-Politifact dataset, which outperforms the baseline model provided in the paper by 37.46\% \cite{rashkin-etal-2017-truth}. Our model performs the best in the Rashkin-Newsfiles dataset in discriminating between Satire and Hoax-Propaganda, achieving an accuracy of 99.43\%, and an F1-Score of 99.71\% in the BERT model. 

For the COVID datasets, the BERT model outperforms the other models. However, current SOTA accuracy on the COVID-AAAI dataset is achieved using a COVID-twitter-BERT model, which was trained on a large corpus of COVID-related tweets, and outperforms our best model by 3.30\% \cite{DBLP:journals/corr/abs-2012-11967}. As explained later, with the cost of adding more in-domain data, the performance tends to improve. Thus, we may have achieved a better score if we would have trained on more in-domain data. 

For the ENRON email spam dataset, our ensemble model performs the best amongst all the models. It outperforms the SOTA hybrid network for spam email detection by 3.70\% in F1-Score \cite{HybridENRON}. Our ensemble model achieves the best accuracy in the SMS Spam dataset, which outperforms the baseline SOTA of the SVM-based model by 0.78\% \cite{SMS}. 

The soft decision ensemble model does not perform better on five of the ten datasets than the standalone models. As the ensemble model takes an average of the softmax decision of the models, a weak classifier gets an equal weight to a strong classifier, which in turn may hurt the final decision. Further investigation may be undertaken to develop a weighted average ensemble of the models for a more robust classifier.

\begin{figure}
\centering
   \includegraphics[width=7cm, height=8cm]{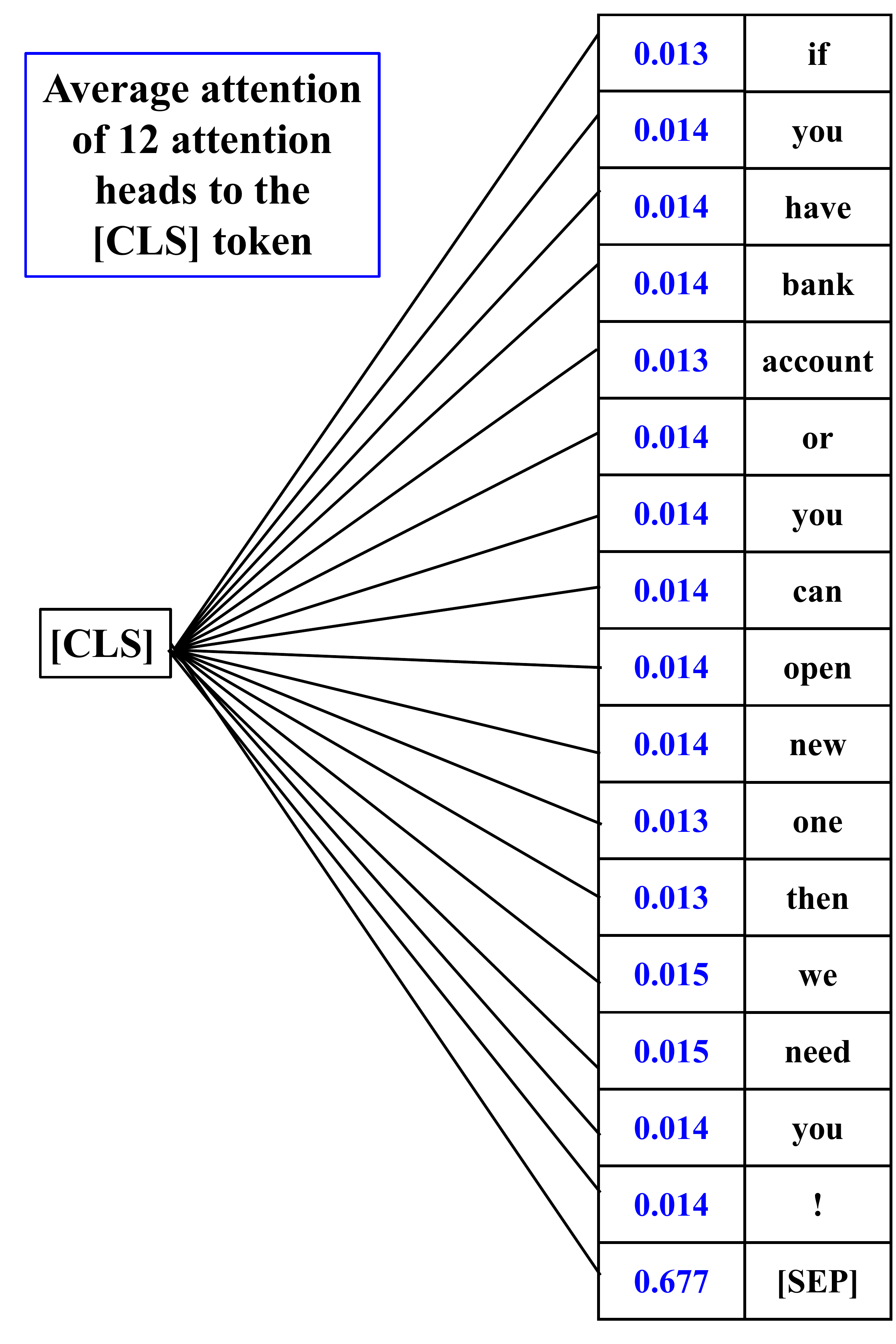}
\caption[Two numerica]{Average attention weight on all attention head of the layer 12 to the $[CLS]$ token for the sentence \textit{``if you have bank account or you can open new one then we need you !''} }
\label{fig:attwt}
\end{figure}

The superior performance of the BERT model comes from the self-attention layer in the transformer blocks, which is also confirmed by Vashishth et al. \cite{vashishth2019attention}. We take the $[CLS]$ token output as the feature vector, and thus the attention heads in each layer for the $[CLS]$ token may have an important impact. We randomly take the deceptive text ``\textit{if you have bank account or you can open new one then we need you !}'' into account and visualize the attention weights of $[CLS]$ token. Figure \ref{fig:attwt} depicts the average attention weights of all attention heads in the final layer. Due to the averaging effect, apart from the $[SEP]$ token, all the words show a close attention weight. However, different attention head focuses on a different part of the text, e.g., we observe that last two attention heads focus on the words `have', `account', `open', `need', and `!', while the third head focuses on the word ``you'', and ``bank''. With a complex mechanism of self-attention and feed-forward network, the BERT model represents the sentence as a 768-D vector, which is used for the downstream deception-detection task. 

\begin{figure}
\centering
  \includegraphics[width=7.5cm, height=6cm]{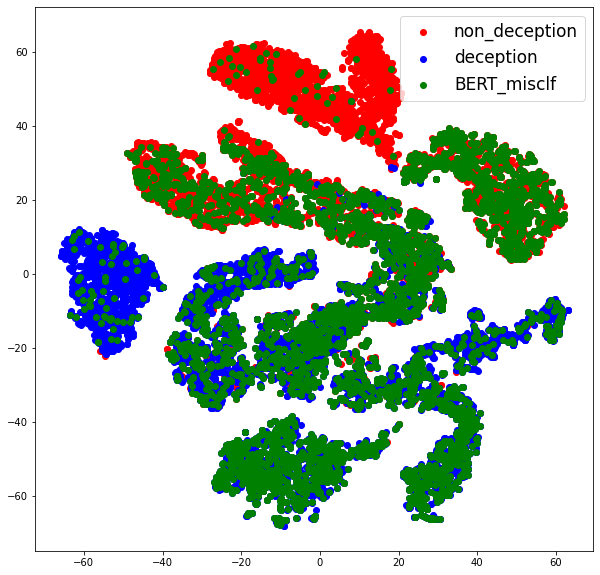}
\caption[Two numerical solutions]{t-SNE embedding (BERT only) for all test data with misclassification.}
\label{fig:tsneBERT}
\end{figure}


We analyze the misclassified samples by the best performing standalone model--BERT. In Figure \ref{fig:tsneBERT}, we plot the BERT embeddings using t-SNE. We observe that BERT does not perform well when deceptive and non-deceptive text has overlapping embedding, which indicates a limitation of the textual feature representation by the BERT model. Future research might be undertaken to develop a variant of the BERT model that can better distinguish between deceptive and non-deceptive text.  

To further analyze the type of misclassified texts, we randomly select False Positive examples (Not deceptive but predicted to be deceptive).
\begin{enumerate}
\item Melania Trump Settles With Daily Mail Parent Over Escort Story
\item Says the cascading effects of climate change contributed to the rise of ISIS
\item The CDC issued its first warning on Jan 8. Trump held campaign rallies on Jan 9, Jan 14, Jan 28, Jan 30, Feb 10, Feb 19, Feb 20, Feb 21, \& Feb 28. He golfed on Jan 18, Jan 19, Feb 1, Feb 15, Mar 7, Mar 8. The first time he admitted the coronavirus might be a problem was Mar 13
\item Justin Theroux Keeps Getting Confused For Justin Trudeau
\item SNL takes a jab at Donald Trump, who doesn't like it kate mckinnon (left) and alec baldwin (right) as clinton and trump nbc universal “saturday night live” takes swings at all political candidates, regardless of party. and with alec baldwin and kate mckinnon portraying donald trump and hillary clinton, they keep hitting the mark.
\end{enumerate}

We observe that all the models have a higher tendency to label a text as deceptive when names of certain political figures are associated with it. For instance, we have the overall False Positive rate (FPR) of 5.31\%. However, the example with the name ``Trump'' has a false positive rate (FPR) of 23.17\%,  ``Obama'' has an FPR of 16.18\%. On the contrary, the non-political names like ``Gates'' have an FPR of 2.01\%. These findings suggest that the models may suffer from bias towards political names.

Next, we analyze the True Negatives (deceptive, but all our models predicted it to be non-deceptive). We randomly select the following samples:
\begin{enumerate}
    \item You have received your mobile content. Enjoy
    \item Celebrities slam Trump decision to end DACA as 'callous,' 'disgusting,' and a 'grave mistake'
    \item Ive been here almost every day.
    \item Forty-five percent of doctors say theyll quit if health care reform passes
    \item Says 57 percent of federal spending goes to the military and just 1 percent goes to food and agriculture, including food stamps

\end{enumerate}

From the True-Negative samples, we observe a wide variety of examples that were misclassified to be non-deceptive. For the first sample, the models probably expect more persuasion to detect the deception. The third sample is a statement by the Missouri governor which was a lie. We infer that it may be hard for any model to detect a text as deceptive without proper context.  These findings raise intriguing questions regarding the extent of the deceptive text, and for the model to successfully detect deception, maybe the context should be a part of the text.


Therefore,based on the overall analysis, the results of the holistic deception detection task supports \textbf{H1}.

\subsection{New-event Deception Detection}
\begin{figure}
\centering
   \includegraphics[width=8cm, height=6cm]{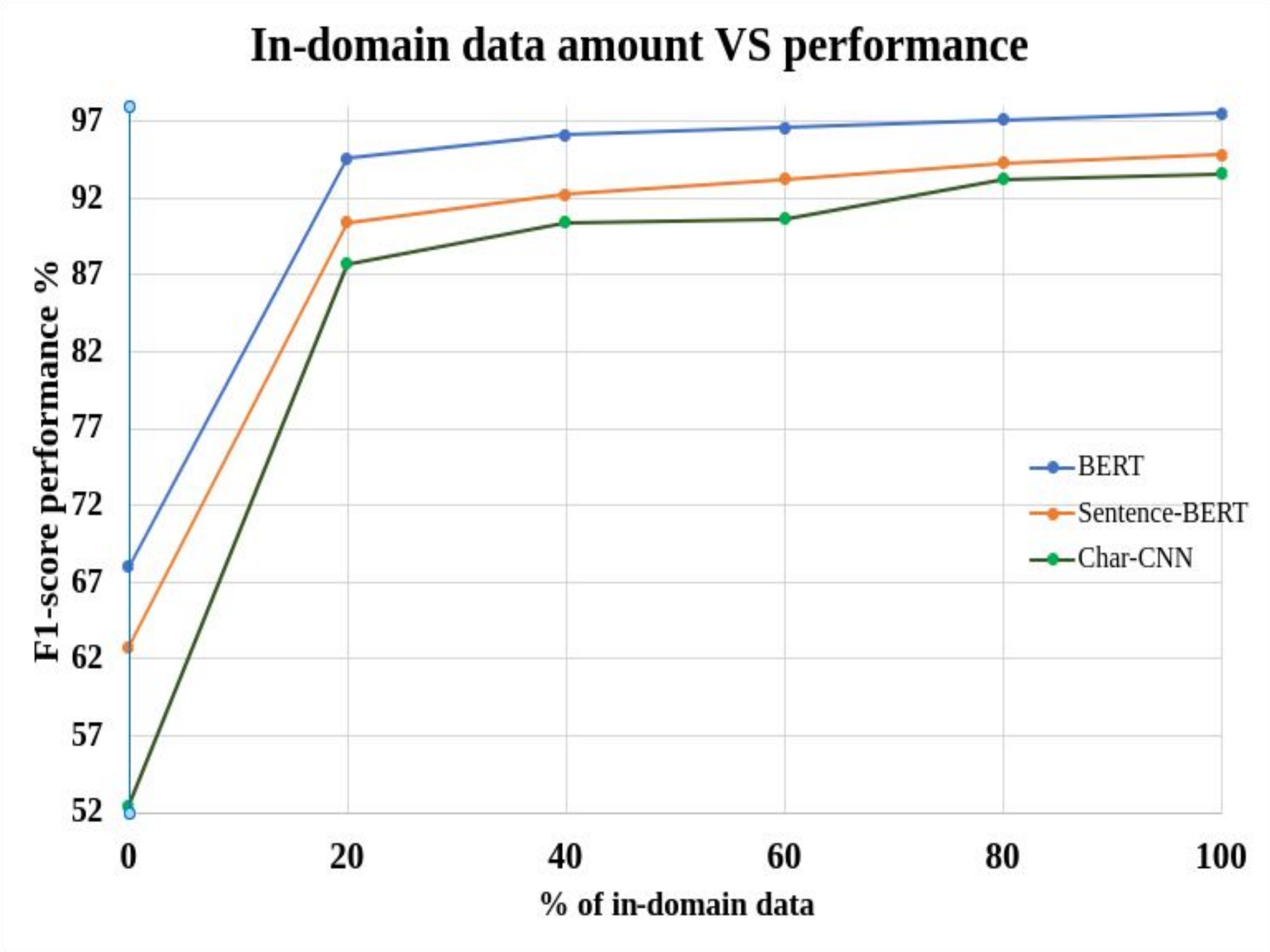}
\caption[Two numerical solutions]{The performance graph of COVID-19 dataset while used different proportion of in-domain data. We observe a cold-start when no in-domain data is present. With the addition of only 20\% in-domain data, the performance improved significantly.}
\label{fig:newevent}
\end{figure}

The holistic model with a complete set of out-of-domain data and a fraction of in-domain data can perform well enough to detect the deceptive text on a new event like the COVID-19 pandemic. From Figure \ref{fig:newevent}, we observe that BERT, SBERT, and Char-CNN give F1-Score performance of 67.96\%, 62.70\%, and 52.39\% respectively while having no knowledge on the COVID event, which indicates a cold-start \cite{1423975}. However, when added with only 20\% in-domain COVID training data, the performance improves sharply to 94.50\%, 90.38\%, and 87.69\%, with an average improvement of 29.84\%. From that point, we gradually increase the in-domain training data by 20\%, and we find the optimal performance by adding 100\% in-domain data. Nevertheless, our holistic model achieves 95.40\% of optimal performance by only seeing the 20\% in-domain training data. Thus,  our results support \textbf{H2}. 

\section{Conclusion and Future Work}

This research presents the holistic deception detection technique where we intend to find a domain-independent system to detect deception. Our general holistic approach outperforms some of the benchmark datasets for deception detection, where we observe the superior performance of the BERT model. Additionally, we analyze the strength of our holistic approach in case of a new event like COVID-19. We show that an out-of-domain general training set with a small fraction of the in-domain training set can help achieve satisfactory performance. Based on our work, there can be several directions for further research --
\begin{itemize}
    \item The BERT and SBERT model work for 512 tokens only. For deception within the long text, models like Longformer, DocBERT can be used \cite{beltagy2020longformer, adhikari2019docbert}.
    \item It is not clear which part of the text may contain the cues to be a deceptive one. Thus, researchers can investigate to localize deception.
    \item The current pre-trained models are not free of bias, which we also observe in this work. Further research can be done to avoid the bias.
    \item The analysis of how deception occurs within the text is still not a clearly studied area. Thus, deceptive text generation can unravel many unexplored areas. Besides, we can investigate certain psycho-linguistic traits like fear, greed, persuasion within the text and quantify these attributes for a stronger holistic deception detection model.
\end{itemize}
\section*{Acknowledgments}

Research was supported in part by grants NSF 1838147, NSF 1838145, ARO W911NF-20-1-0254.
The views and conclusions contained in this document are those of the authors and not of
the sponsors. The U.S. Government is authorized to reproduce and distribute reprints for
Government purposes notwithstanding any copyright notation herein.

\bibliographystyle{acl_natbib}
\bibliography{ranlp2021}


\end{document}